\documentstyle[amssymb,aps,preprint]{revtex}

\begin{document}
\title{Ga-NMR local susceptibility of the kagom\'{e}-based magnet SrCr$_{9p}$Ga$%
_{12-9p}$O$_{19}.$ A high temperature study.}
\author{L.Limot$^{1}$, P.Mendels$^{1},$ G.Collin$^{2}$, C.Mondelli$^{3}$, H.Mutka$%
^{3}$, and N.Blanchard$^{1}$}
\address{$^{1}$Laboratoire de Physique des Solides, UMR 8502, Universit\'{e} de
Paris-Sud, 91405 Orsay, France\\
$^{2}$Laboratoire L\'{e}on Brillouin, CE Saclay, CEA-CNRS, 91191,
Gif-sur-Yvette, France\\
$^{3}$Institut Laue-Langevin, B.P. 156, F-38042 Grenoble Cedex 9, France}
\maketitle

\begin{abstract}
We report a high-$T$ Ga-NMR study in the kagom\'{e}-based antiferromagnetic
compound SrCr$_{9p}$Ga$_{12-9p}$O$_{19}$ ($.81\leq p\leq .96$), and present
a refined mean-field analysis of the local NMR susceptibility of Cr
frustrated moments. We find that the intralayer kagom\'{e} coupling is $%
J=86(6)$ K, and the interlayer coupling trough non-kagom\'{e} Cr moments is $%
J^{\prime }=69(7)$ K. The $J^{\prime }/J=0.80(1)$ ratio confirms the common
belief that the frustrated entity is a pyrochlore slab.
\end{abstract}

\newpage

The kagom\'{e}-based compound SrCr$_{9p}$Ga$_{12-9p}$O$_{19}[$SCGO$(p),p<1]$
has been receiving considerable attention as a model system for
geometrically frustrated physics \cite{SCGOgen}. Magnetic properties of SCGO$%
(p)$ arise from Heisenberg Cr$^{3+}$ ions $(S=\frac{3}{2})$. Despite
macroscopic susceptibility $(\chi _{macro})$ measurements which indicate
that a strong antiferromagnetic interaction couples neighboring spins ($%
\Theta _{macro}\approx 500-600$ K), spin-spin correlation never exceeds
twice the Cr$-$Cr distance \cite{Neutron}. No classical N\'{e}el long range
order occurs. On the contrary, SCGO$(p)$ continues to display Curie-Weiss
behavior well below $\Theta _{macro}$ as suggested by the linear $T$%
-dependence of $\chi _{macro}^{-1}$, extending mean-field predictions to an
unusual temperature domain $(T<<\Theta _{macro})$.

Geometric frustration arises from interacting spins on kagom\'{e} networks.
However, the crystal structure is not pure kagom\'{e} since Cr$^{3+}$ ions
occupy both kagom\'{e} $[$Cr$(12k)]$ and non-kagom\'{e} $[$Cr$(2a),$Cr$%
(4f_{vi})]$ sites, as shown in the inset of Fig.\ref{fig.1}. Cr$(2a)$ sites
link two kagom\'{e} layers. The kagom\'{e}-Cr$(2a)$-kagom\'{e} slabs are
isolated from one another by Cr$(4f_{vi})-$Cr$(4f_{vi})$ pairs which Lee 
{\it et al. }showed to form spin singlets with a binding energy of $\Delta
=216$ K \cite{Lee}. They also suggested that the kagom\'{e}-Cr$(2a)$%
-kagom\'{e} slab may be regarded as a quasi-2D highly frustrated system,
which they named pyrochlore slab$.$ However there is a lack of experimental
evidence concerning couplings which ensure this\ slab structure. Coupling of
the two kagom\'{e} planes through Cr$(2a)$ is not known, nor is the
intralayer kagom\'{e} coupling. One way to shed light into this problem,
would be a refined high-$T$ analysis of the magnetic susceptibility of Cr
frustrated moments. Unfortunately this is prevented in $\chi _{macro}$ by
the fact that there are contributions from all magnetic sites, frustrated
and non-frustrated. Moreover, $\chi _{macro}$ is also sensitive to disorder
as non-magnetic Ga$^{3+}$ ions are always present on Cr$^{3+}$ sites (the $%
p=1$ compound is not stable) \cite{Schiffer}. As we show here, such an
analysis can be done by $^{69}$Ga and $^{71}$Ga $(I=\frac{3}{2})$ NMR since
Ga nuclei on site $4f$ directly probe both Cr$(12k)$ and Cr$(2a)$ moments
(see. inset Fig.\ref{fig.1}). By varying Ga/Cr substitution from 4\% to 19\% 
$(p=.81,.89,.96)$, we are able to separate and identify on a firm
experimental ground the substitution-related susceptibility from the
intrinsic frustrated susceptibility of the slab \cite{Mendels,Limot}. We
present here high-$T$ results and mean-field analysis of the slab
susceptibility, which enables us to evaluate $J^{\prime }$ and $J$,
respectively the Cr$(2a)-$Cr$(12k)$ coupling and the Cr$(12k)-$Cr$(12k)$
coupling. We find $J^{\prime }=69(7)$ K and $J=86(6)$ K. The $J^{\prime
}/J=0.80(1)$ ratio confirms the common belief that the frustrated entity is
a pyrochlore slab.

All samples are ceramics and were synthesized by solid state reaction as
detailed elsewhere \cite{Limot}. Reaction products were checked by standard
x-ray diffraction and refined by neutron diffraction (see Table I). $\chi
_{macro}$ yielded results in agreement with previous concentrations reported
in literature ($\Theta _{macro}=439(22),501(14),608(14)$ K for $%
p=.81,.89,.96 $ respectively) \cite{SusMacro}.

An extended NMR analysis of the spectrum was presented in previous papers 
\cite{Limot,Keren}. Three different Ga NMR sites can be resolved: Ga$(4e)$,
Ga$(4f)$ and Ga substituted on Cr sites $[$Ga$(sub.)]$. Ga$(4f)$ is at the
heart of frustrated physics (inset Fig.\ref{fig.1}). It is exclusively
coupled to Cr$(12k)$ and Cr$(2a)$ by a Ga$-$O$-$Cr hyperfine interaction,
and is the object of the present study.

A set of spectra ranging from 150 K to 410 K were performed in a $\sim $7
Tesla permanent magnet. We used a $\pi /2$, $\pi $ sequence with spacing $%
\tau =18$ $\mu s$ between pulses, and as the applied frequency was swept
over a $\sim $1 MHz range, with a step interval of 75 kHz, spin echo signals
were recorded. The overall spectrum was then obtained by summing all the
spin echo Fourier transforms \cite{Clark}. Another set of spectra ranging
from 15 K to 200 K were obtained in a sweep field setup, i.e. by varying an
external magnetic field at a constant applied frequency $40.454$ MHz, and
recording the integral of the spin echo signal. Again we used a conventional 
$\pi /2$, $\pi $ sequence, but with spacing $\tau =28$ $\mu s$.

A typical set of high-$T$ ($T\geq 100$ K) field sweep spectra for the $p=.96$
sample, recorded around 3 Tesla, is reported in Fig.\ref{fig.1}. In this
field window two lines are unambiguously resolved. Ga$(4f)$ is the main line
and the hump at its right is Ga$(sub.)$. The negligible contribution of
Ga(sub.) to the spectrum reflects the nearly stoichiometric value of the
sample. The sharp non-symmetric features of the Ga$(4f)$ line originate from 
$T$-independent quadrupole effects ($^{71}\nu _{Q}(4f)\simeq 2.9$ MHz, $\eta
\simeq 0)$ related to local charge environment around the nucleus. Upon
cooling the NMR line shifts to the left with respect to its zero reference
position, without any appreciable broadening. The $T$-dependence of the
shift $(K)$ stems from the Ga$(4f)$ magnetic environment. $K$ provides a
direct measurement of the magnetic susceptibility of the Cr$(12k)$ and Cr$%
(2a)$ ions (to be detailed below).

The temperature dependence of $K^{-1}$ is presented in Fig.\ref{fig.2} for
all samples. The high-$T$ data ($T\gtrsim 100$ K$)$ follows a Curie-Weiss
law as suggested by the linear variation of $K^{-1}$. By extrapolating $%
K^{-1}$ to $0$ we extract the NMR Curie-Weiss temperature $(\Theta _{NMR}).$
We evaluate it to be $453(50)$ K, $469(44)$ K, and $484(40)$ K for $%
p=.84,.89,.96$ respectively, of the same order of $\Theta _{macro},$ but
somewhat lower. The increase in $\Theta _{NMR}$ with $p$ reflects a better
lattice coverage as one would expect from mean-field theory. Here again
Curie-Weiss behavior extends well below $\Theta _{NMR}.$ Further lowering $T$%
, $K$ in all samples eventually deviates from Curie-Weiss behavior,
flattening and even decreasing below $T=40-50$ K. We address the reader to
other publications for a detailed discussion on the low-$T$ behavior of $K$ 
\cite{Mendels,Limot}.

We now turn to discuss the high-$T$ behavior of $K$ within a mean-field
analysis. First we would like to point out that Cr$(12k)$ and Cr$(2a)$ do
not have same magnetic environment. Cr$(2a)$ has 6 nearest neighbors of Cr$%
(12k),$ whereas Cr$(12k)$ has 5 nearest neighbors (4 Cr$(12k)$ and 1 Cr$(2a)$%
). Furthermore Cr$(12k)-$Cr$(12k)$ coupling $(J)$ and Cr$(12k)-$Cr$(2a)$
coupling $(J^{\prime })$ are a priori not the same. In these regards our
mean-field analysis is similar to that performed in ferrimagnetic compounds
which bear inequivalent magnetic sites.

To justify this analysis we must argue how $J$ and $J^{\prime }$ are the
only relevant couplings in the slab. A study of a series of chromium oxides
indicates that the Cr-Cr interaction is dominated by direct exchange \cite
{Motida}. It was shown that such an interaction varies rapidly with Cr$-$Cr
distance (and Cr$-$O$-$Cr angle), being $\approx 0$ when $d_{Cr-Cr}\gtrsim
3.1$ \AA . Therefore only interactions between Cr$(12k)-$Cr$(12k)$ ($%
d_{Cr-Cr}=2.895$ \AA ) and Cr$(12k)-$Cr$(2a)$ ($d_{Cr-Cr}=2.971$ \AA ) are
relevant in the frustrated unit, with $0<J^{\prime }<J$. Cr$(2a)-$Cr$(2a)$
interaction is $\approx 0$ as $d_{Cr-Cr}=5.80$ \AA . Note that the
kagom\'{e} layer is distorted, therefore $J$ represents the average coupling
between Cr$(12k)$.

On the basis of these considerations, the frustrated spin Hamiltonian is 
\[
{\cal H}=J\sum_{\left\langle i,j\right\rangle }\left[ (\overrightarrow{S}%
_{12k})_{i}\cdot (\overrightarrow{S}_{12k}{\bf )}_{j}+\varepsilon (%
\overrightarrow{S}_{12k})_{i}\cdot (\overrightarrow{S}_{2a})_{j}\right] , 
\]
where we have introduced $0<\varepsilon $ $=J^{\prime }/J<1.$ The first term
is the isotropic Heisenberg interaction between spins $\overrightarrow{S}%
_{12k}$ on the kagom\'{e} layer. The second term is the interaction between $%
\overrightarrow{S}_{12k}$ and $\overrightarrow{S}_{2a}$ spins. The summation
is taken over neighboring spins of the slab.

Using a mean-field approach the $\chi _{12k}$ and $\chi _{2a}$
susceptibilities per site of Cr$(12k)$ and Cr$(2a)$ can be written 
\begin{equation}
\chi _{12k}=\frac{\mu _{eff}^{2}}{3k_{B}Tf(T)}(1-\frac{p_{2a}S(S+1)%
\varepsilon J}{3k_{B}T}),  \label{1}
\end{equation}
and 
\begin{equation}
\chi _{2a}=\frac{\mu _{eff}^{2}}{3k_{B}Tf(T)}(1+\frac{6p_{12k}S(S+1)(2/3-%
\varepsilon )J}{3k_{B}T}),  \label{2}
\end{equation}
where $f(T)=1+4p_{12k}S(S+1)J/3k_{B}T-6p_{12k}p_{2a}(\varepsilon
JS(S+1)/3k_{B}T)^{2}.$ The two expressions are different and reflect the
fact that the two sites have different magnetic environments. The effective
magnetic moment is $\mu _{eff}=3.83\mu _{B}$, typical of $S=3/2$ Cr$^{3+}$
ions \cite{Ohta}. $p_{12k},p_{2a}$ are the occupation of the Cr$(12k)$ and Cr%
$(2a)$ sites known from neutron refinement (cf. Table I). The only unknown
parameters are $J$ and $\varepsilon $. In Fig.\ref{fig.3} we show the
evolution of $\chi _{12k}$ and $\chi _{2a}$ when increasing $\varepsilon $,
with a fixed value $J.$ In the limit $\varepsilon =0,$ $\chi _{12k}$ follows
a Curie-Weiss law with a kagom\'{e} Curie-Weiss temperature $%
4p_{12k}JS(S+1)/3.$ Whereas $\chi _{2a}$ follows a Curie law as we have
freed the Cr$(2a)$ spin from its magnetic environment. As $\varepsilon $ is
increased, the pure paramagnetic behavior of $\chi _{2a}$ progressively dies
out, while $\chi _{12k}$ is little affected. Of interest is when the
kagom\'{e}-Cr$(2a)$-kagom\'{e} slab behaves as a pyrochlore slab. By
pyrochlore slab we mean that both magnetic sites are identical $\chi
_{12k}=\chi _{2a},$ i.e. $\varepsilon =\varepsilon
_{0}=4p_{12k}/(6p_{12k}-p_{2a}).$ For an homogeneous substitution $%
(p_{12k}=p_{2a}=p)$ $\varepsilon _{0}=0.8,$ reflecting the fact that the
effective exchange field on Cr$(2a)$ $(6J^{\prime })$ and on Cr$(12k)$ $%
(4J+J^{\prime })$ are then identical. We stress that this relation is
independent on distortion on the kagom\'{e} layer, as it only involves the
average exchange constant $J.$

We may express the NMR shift $K$ using the mean-field expressions (\ref{1})
and (\ref{2}) 
\begin{equation}
K\propto (9p_{12k}\chi _{12k}+3\delta p_{2a}\chi _{2a})  \label{3}
\end{equation}
where we have introduced $\delta =A^{\prime }/A.$ $A,A^{\prime }$ are the
hyperfine constants which are respectively a measurement of the Cr$(12k),$Cr$%
(2a)$ moment at the nuclear site, and just influence the thermal behavior of 
$K$ by their ratio $\delta .$ The two terms of expression (\ref{3}) are
multiplied by the number of neighboring Cr$(12k)$ and Cr$(2a)$ linked by the
oxygen ions to the Ga$(4f)$ nucleus, respectively $9$ and $3$. We can fix a
lower bond to $\delta $ by some simple structural remarks. The hyperfine
bridge Ga$(4f)-$O$-$Cr$(12k)$ yields bond lengths of $d_{Ga(4f)-O}=1.866$
\AA\ and $d_{O-Cr(12k)}=2.052$ \AA . The hyperfine bridge Ga$(4f)-$O$-$Cr$%
(2a)$ yields bond lengths of $d_{Ga(4f)-O}=1.866$ \AA\ and $%
d_{O-Cr(2a)}=1.971$ \AA . The small difference in length and the similarity
for angles between these bonds leads to a stronger coupling to Cr$(2a)$ then
to Cr$(12k),$ and $\delta >1.$ A more precise calculation suggests $\delta
\approx 2$ \cite{Keren}. It is instructive to compare $K$ to the expression
of the susceptibility of the kagom\'{e}-Cr$(2a)$-kagom\'{e} slab that one
would obtain from a macroscopic measurement 
\[
\chi _{slab}=6p_{12k}\chi _{12k}+p_{2a}\chi _{2a} 
\]
where $\chi _{12k}$ and $\chi _{2a}$ are now weighted by $6$ and $1$,
respectively the number of Cr$(12k)$ and Cr$(2a)$ per formula unit of SCGO$.$
$K$ is an independent measurement of the slab susceptibility, but with a
different weighting than $\chi _{slab}$. Using $\delta \approx 2,$ we see
that $\chi _{2a},\chi _{12k}$ impact $K$ in a ratio of $6/9$ instead of $1/6$
for $\chi _{slab}.$ $K$ on the contrary of $\chi _{slab}$ is very sensitive
to temperature dependence of $\chi _{2a}$, meaning that it is also very
sensitive to $\varepsilon .$

The shift $K$ reseals three unknown parameters $J,$ $\varepsilon ,$ and $%
\delta $. Our aim is to determine their values by fitting for $T\geq 100$ K
the shift-data presented in Fig.\ref{fig.2}. There are an infinity of $%
(\varepsilon ,J,\delta )$-set of values which reproduce the linear variation
of $K^{-1}$ with $T$. In Fig.\ref{fig.4} we present the set of $(\varepsilon
,J,\delta )$-values obtained when fitting $K^{-1}$ of the $p=.96$ sample
with expression (\ref{3}). As shown, for a given value of $\varepsilon $
(abscissa) there is an associated value $\delta $ (upper panel) and $J$
(bottom panel)$.$ In order to seize the evolution of this set of parameters
it is instructive to examine the dependence of $\delta $ on $\varepsilon .$
In the limit $\varepsilon \rightarrow 0$, we see that $\delta \rightarrow 0.$
In fact when $\varepsilon =0$, $\chi _{2a}$ follows a Curie law. Therefore
to reproduce the linear variation of $K^{-1}$ there is only one possibility:
uncouple the Ga$(4f)$ nuclei from Cr$(2a)$, setting $\delta =0.$ As $%
\varepsilon $ is increased, $\delta $ monotonically increases since Curie
contribution from $\chi _{2a}$ dies out. But when $\varepsilon =\varepsilon
_{0}$ expression (\ref{3}) reduces to $K\propto (9p_{12k}+3\delta
p_{2a})\chi _{12k}$, since in this case $\chi _{12k}=\chi _{2a}=\chi
_{slab}/(6p_{12k}+p_{2a}).$ The $T$-dependence of the shift is then
completely determined by the ($\varepsilon ,J)$ pair of values, whatever the
value of $\delta $. When $\varepsilon >\varepsilon _{0}$, we recover the
monotonic increase in $\delta .$ Of interest is also the slow monotonic
decrease of $J$ with increasing $\varepsilon .$ Since $0<J^{\prime }<J$,
i.e. $0<\varepsilon <1,$ we conclude that $80$ K$\lesssim J\lesssim 100$ K.
However only the set of values for which $\delta >1$ bear a physical
solution. Two possibilities subsist: (i) the set of values close to $%
\varepsilon \approx \varepsilon _{0}$ (ii) the set of values for $%
\varepsilon \gtrsim 4.$ Case (ii) is a non-physical solution. This leads us
to one of our major findings: the high-$T$\ linear behavior of $K^{-1}$\
reflects the physics of a pyrochlore slab, with the consequence that $%
K\propto \chi _{slab}.$

Results for all samples are presented in Table I. Errors are governed by
error bars on $p_{12k}$ and $p_{2a}$ from neutron diffraction data. Coupling
constants are found independent on lattice coverage given the negligeable
variation of lattice parameters with substitution. Turning to the value of $%
J,$ it is very satisfying to find a coupling constant close to the one
observed in Cr$_{2}$O$_{3}$, where Cr$^{3+}$ ions have a local octahedral
environment ($d_{Cr-Cr}=2.890$ \AA , Cr$-$O$-$Cr$=93.1%
{{}^\circ}%
$) similar to Cr$(12k)$ ($d_{Cr-Cr}=2.895$ \AA , Cr$-$O$-$Cr$=93.8%
{{}^\circ}%
)$. There, it was established that the exchange coupling between neighboring
Cr is $77(3)$ K\cite{Samuelsen}. We also performed a consistency test by
calculating the expected values of $\Theta _{NMR}$ using results reported in
Table I. The analytical expression of $\Theta _{NMR}$ is obtained from the
high-$T$ limit $(T>>S(S+1)J)$ of expression (\ref{3})$.$ We find $%
456(56),460(33)$ and $495(35)$ K for $p=.81,.89$ and $.96$ respectively, in
agreement with experimental values of $\Theta _{NMR}.$ Finally by using the
susceptibility of Cr$(4f_{vi})-$Cr$(4f_{vi})$ pairs \cite{Lee}, we are able
to reproduce the high-$T$ $\chi _{macro}-$data where the susceptibility $%
\chi _{slab}$ adds on to the Cr$(4f_{vi})$ and to the defect induced
susceptibilities.

In conclusion, we have evidenced that the Curie--Weiss behavior of the
frustrated slab susceptibility extends to much lower temperatures than the
average Cr$-$Cr interaction. A high-$T$ mean-field analysis allowed us to
determine couplings involved in the frustrated physics. We confirm the
common belief that frustration in SCGO$(p)$ arises from a pyrochlore slab.

We wish to acknowledge fruitful discussions with A.\ Keren, C. Lhuiller, F.
Mila, M. Horvati\'{c}, and B. Du\c{c}ot.

\newpage

\begin{center}
$
\begin{tabular}{cccccc}
\hline\hline
$p$ & $p_{12k}$ & $p_{2a}$ & $p_{4f_{vi}}$ & $J$ (K) & $\varepsilon $ \\ 
\hline\hline
$.81(1)$ & $.79(1)$ & $.95(2)$ & $.81(1)$ & $94(9)$ & $.82(2)$ \\ \hline
$.89(1)$ & $.89(1)$ & $.94(2)$ & $.86(1)$ & $85(6)$ & $.81(2)$ \\ \hline
$.96(1)$ & $.96(1)$ & $1.00(2)$ & $.94(1)$ & $86(6)$ & $0.80(1)$ \\ 
\hline\hline
\end{tabular}
$
\end{center}

TABLE I. Nominal occupation of Cr sites and occupations of Cr$(12k)$, Cr$%
(2a),$ and Cr$(4f_{vi})$ sites from neutron refinement. $J$ and $\varepsilon 
$ are the couplings extracted from the high-$T$ mean-field analysis for the
three samples.

\begin{figure}[tbp]
\caption{Typical $^{71}$Ga field sweep obtained at 40.454 MHz for $p=.96$.
>From right to left: 150 K, 120 K, 100 K. The zero shift reference position
is 3.116 Tesla. Horizontal arrow indicates shift direction upon cooling.
Inset: Crystal structure of ideal SrCr$_{9}$Ga$_{3}$O$_{19}$. The thick
dashed lines show the hyperfine coupling paths of Ga($4f$) nucleus to
frustrated Cr moments. Light grey circles are oxygens. Cr$(12k)$ are
arranged to form a distorted kagom\'{e} network and are coupled through Cr$%
(2a)$. The full structure is Cr$(4f_{vi})-$Cr$(4f_{vi})$/kagom\'{e}-Cr$(2a)$%
-kagom\'{e}/Cr$(4f_{vi})-$Cr$(4f_{vi})$, etc.}
\label{fig.1}
\end{figure}

\begin{figure}[tbp]
\caption{$K^{-1}$ versus T down to 15 K, for all samples. Minor second-order
quadrupole corrections have been performed. The straight line extrapolation
of $K^{-1}=0$ yields $\Theta _{NMR}$.}
\label{fig.2}
\end{figure}

\begin{figure}[tbp]
\caption{Calculated mean-field susceptibilities $\protect\chi _{12k}$ and $%
\protect\chi _{2a}$ of the Cr$(12k)$ and Cr$(2a)$ moments versus $T$, with
increasing Cr$(2a)-$Cr$(12k)$ coupling$.$ $J$ is fixed to $85$ K and $%
p_{12k}=p_{2a}=1.$}
\label{fig.3}
\end{figure}

\begin{figure}[tbp]
\caption{The $(\protect\varepsilon ,J,\protect\delta )$ set of values which
reproduce the linear variation of $K^{-1}$ with $T$ of the $p=.96$ sample.
For a given value of $\protect\varepsilon $ (abscissa) there is an
associated value $\protect\delta $ (upper panel) and $J$ (bottom panel).}
\label{fig.4}
\end{figure}

\end{document}